\documentclass[12 pt]{article}
\usepackage[]{psfig}
\usepackage[american]{babel}
\usepackage{txfonts}
\usepackage{stmaryrd}
\usepackage{geometry}
\def\fh{\hbox{$.\!\!^{\rm h}$}}
\def\fm{\hbox{$.\!\!^{\rm m}$}}
\def\farcs{\hbox{$.\!\!^{\prime\prime}$}}
\usepackage{setspace}

\begin{document}
\begin{large}

\begin{center}{\bf CCD photometry of 23 minor planets}
\vskip2cm
P. Sz\'ekely$^{1,3}$, L. L. Kiss$^{1,2,3,*}$, Gy. M. Szab\'o$^{1,3}$,
K. S\'arneczky$^{3}$\\
B. Cs\'ak$^{1,3}$, M. V\'aradi$^{1,3}$, Sz. M\'esz\'aros$^{1,3}$
\end{center}
\vskip2cm
1: Department of Experimental Physics \& Astronomical Observatory, 
SZTE University, H-6720 Szeged, D\'om t\'er 9., Hungary\\
2: School of Physics A28, University of Sydney, NSW 2006, Australia\\
3: Guest observer at Konkoly Observatory, Hungary\\

\vskip2cm

\noindent Send offprint requests and proofs to corresponding author: P. Sz\'ekely:\\
\vskip5mm
\noindent Department of Experimental Physics\\ 
SZTE University, H-6720 Szeged, D\'om t\'er 9., Hungary\\
\vskip5mm
\noindent E-mail: pierre@physx.u-szeged.hu, tel: +36 62/544-668, fax: +36 62/420-154\\
\vskip5mm
\noindent *: On leave from University of Szeged, Hungary
\clearpage

{\bf Abstract:} We present CCD photometric observations 
of 23 main-belt asteroids, of which 8 have never been observed before thus the data of these objects are the first in the literature. The 
majority showed well detectable light variations,
exceeding 0\fm1. We have determined synodic periods for  756 Lilliana (9\fh36),1270 Datura (3\fh4), 1400 Tirela (13\fh36), 1503 Kuopio (9\fh98), 3682 Welther (3\fh59), 7505 Furushu (4\fh14) and 11436 1969 QR (2\fh6), while uncertain period estimates were possible for 469 Argentina (12\fh3), 546 Herodias (10\fh4) and 1026 Ingrid (5\fh3). The shape of the lightcurves of 3682 Welther changed on a short time-scale and showed dimmings that might be 
attributed to eclipses in a binary system.
For the remaining objects, only lower limits of the periods and amplitudes 
were concluded.

\clearpage

{\bf Introduction}
\vskip1cm
More than 200 years after the discovery of the first main-belt asteroid, 1
Ceres, highly automatized and sensitive surveys yielded an unimaginable number
of newly discovered minor planets. As of writing this paper, the numbered
asteroids exceed 65.000, while the total number of asteroid  discoveries
reached 220.000\footnote{\tt 
http://cfa-www.harvard.edu/iau/lists/ArchiveStatistics.html}. The process is 
still accelerating, and the observations arrived to the realm of the faintest, 
thus the smallest  objects (e.g. Morbidelli et al. 2002). On the other hand, 
the huge number of objects causes a strong
limitation in determining and understanding their physical properties. The
overwhelming majority of these newly found objects goes without further notice
and a large body of solar system objects is accumulating without known 
observable properties (e.g. rotation, shape). That is why a new field has
emerged, to which even well-equipped amateur astronomers or small college
observatories can make significant contribution: follow-up photometric
observations of moderately faint and small asteroids (roughly from
12 mag to 17 mag). The study of asteroid spin
rates (Binzel et al. 1989, Fulchignoni et al. 1995, Pravec \& Harris 2000), 
requires a large amount of time-resolved
photometric data for the determination of synodic periods (see Angeli et 
al. 2001 for a recent example), so that period determination of minor 
planet lightcurves is a potentially useful project for small and 
moderate-sized
telescopes.

Our group has been working on a photometric project addressing rotational
properties of main-belt and Near-Earth asteroids since 1998 (S\'arneczky et al.
1999, Kiss et al. 1999, Szab\'o et al. 2001). The main aim of this project
is to extend available data toward fainter and therefore smaller minor 
planets. 
Here we report the results of new observations of 23 minor planets obtained 
in the years 1999-2003. 

\vskip1cm
{\bf Observations}
\vskip1cm
We carried out CCD observations on 39 nights from September, 1999 to February, 
2003.
Geometric data of the examined asteroids are listed in Table\ \ref{aspect}. 
The full log of observations is given in Table\ \ref{logs}.
Standard Johnson R filtered and unfiltered data were obtained using the 
60/90/180 cm
Schmidt-telescope of the Konkoly Observatory, equipped with a  Photometrics 
AT200 CCD camera
(1536$\times$1024 KAF 1600 MCII coated CCD  chip). The projected area is 
29$^\prime$ $\times$
18$^\prime$, which corresponds to an angular  resolution of 1\farcs1/pixel. 
The operational
temperature of the camera was below --40 $^\circ$C.

The bulk of the data was acquired with the 0.4m Cassegrain-telescope of the 
Szeged Observatory, which is located in the outskirts of the city of Szeged, 
though hampered by heavy light pollution. 
This telescope was used with a cooled SBIG ST--9E CCD camera
(512$\times$512 20$\mu$m sized, 2$\times$2 binned pixels, angular resolution 1\farcs4/pixel). The field of view 
was $6^\prime\times6^\prime$. Most single-filtered $R$-band and unfiltered 
observations were made with this instrument. The achieved photometric accuracy 
varies between 0.01--0.1 mag, depending on the brightness of the target and 
weather conditions, typically 0\fm05. The precision was estimated with the rms scatter of comparison {\it minus} check data.

There were several criteria when selecting an asteroid for photometric 
observations. Because of the
small-sized apertures of these telescopes we were limited to observe 
asteroids usually
brighter than 17\fm0 with the 0.6m Schmidt and 15\fm0 with the 0.4m 
Cassegrain, respectively. 
Exposure times were constrained by two factors. Firstly, objects were not 
allowed to move more
than the half of the full width at half maximum of the stellar profiles 
(varying from night to
night). On the other hand, signal-to-noise ratio had to be kept at least 10. 
This parameter was
estimated by comparing the peak pixel values with the sky background during 
the observations.

The CCD observations were reduced with standard routines in IRAF\footnote
{IRAF is distributed by the National Optical Astronomy Observatories,
which are operated by the
Association of Universities for Research in Astronomy, Inc., under  
cooperative agreement with the National Science Foundation.}, including 
flat-field correction utilizing sky-flat images taken during 
the evening or morning twilight. Differential magnitudes were calculated with 
quick aperture photometry ({\it qphot} and {\it phot} routines) using two or more 
nearby comparison stars of similar brightnesses, selected from the Guide Star 
Catalogue (GSC). Because of the trailed images of asteroids we did not attempt 
point spread function (PSF) photometry. In cases of larger observational 
scatter, we rejected the bad data points and performed a noise filtering of the data by taking median means of 3, 5, 7 or 9 lightcurve points.

Period determination was carried out by a self-developed code, which 
implemented a modified
Phase Dispersion Minimization method (Stellingwerf 1978). The modification 
included the
determination of the nightly photometric zero-points caused by the use of 
different comparison
stars from night to night. The optimal period and magnitude shifts were 
determined by a
grid-search method. Individual data are available electronically at the CDS 
via anonymous ftp to
{\tt cdsarc.u-strasbg.fr (130.79.128.5)} or via\\
{\tt http://cdsweb.u-strasbg.fr/cgi\_bin/qcat?J/A+A/.../...}.

\clearpage
\vskip1cm
{\bf Results}
\vskip1cm

In this section of our paper we present the simple and composite lightcurves of individual asteroids and discuss the obtained rotational periods.

\vskip1cm

{\bf 469 Argentina}: 469 Argentina was the largest observed asteroid by us 
(with a diameter near 130 km). 
There are unequal maxima and noticeable near-linear branches in the first 
third of the presented unfiltered lightcurve. Hazy weather in the second half 
of the night greatly increased the scatter. Although the amplitude is fairly 
low, this asteroid showed well-detectable light variations ranging 0\fm14. The 
lightcurve is plotted in Fig. \ref{argentina}. The estimated synodic period is 12\fh3. The composite lightcurve with use of two first nights is presented in Fig. \ref{argentinacomp}. Wang (2003) gives a period of approximately 3 hours with low reliability for this asteroid, although she did not present lightcurve.

{\bf 531 Zerlina}: During our one night run 531 Zerlina showed apparently monotonic variation over about 4 hours, as can be seen in Fig. \ref{zerlina}. The R-filtered range was over 0\fm17. This small minor 
planet probably has long rotation period. On the web site\footnote{\tt http://obswww.unige.ch/\~{}behrend/page2cou.html} of Raoul Behrend (Observatoire de Gen$\mathrm{\grave{e}}$ve) one can find data for this asteroid. The observer was Bernard Christophe. The given values are 2\fh352 for the synodic period and 0\fm12 for the amplitude of the light variation, respectively. However, judged from the rotational phase diagram found there this applied period does not fit very well.

{\bf 546 Herodias}: Although this mid-sized minor planet had well-detectable 
light variations, the small, 0\fm07 amplitude become indiscernible later, due 
to unfavourable observing circumstances. The best fit yielded a synodical 
period of 10\fh4, however this is a fairly uncertain value.
The unfiltered lightcurves of 546 Herodias are presented in Fig. \ref{herodias}.
We found no other lightcurve in the literature.

{\bf 549 Jessonda}: The short single-night lightcurve of 549 Jessonda did not show a clear pattern of variation above the observational scatter. On the web site of Collaborative Asteroid Lightcurve Link, CALL)\footnote{\tt 
http://www.minorplanetobserver.com/DATA/submissions\_results.htm} one can find rotational parameters for this asteroid. Stephen M. Brincat and Robert Koff. made the observations listed on the CALL web site and Laurent Bernasconi on the Behrend web site, respectively. Based on 5 observing sessions the synodical period is 4\fh5 and the amplitude is 0\fm1, although they publish no lightcurve. The reliability of this data is low. Besides the detected slight amplitude our lightcurve shows data obtained amidst unfavourable weather conditions, therefore it is not inconsistent with the above mentioned results of CALL.

{\bf 697 Galilea}: This asteroid was observed in December, 2001 by Sheridan (2002), who obtained a period of 16\fh538$\pm0\fh002$ and amplitude of 0\fm28.
Our one-night lightcurve of 1.7 hours duration shows a moderate linear trend consistent with this period and amplitude, as can be seen in Fig. \ref{galilea}.

{\bf 756 Lilliana}: 756 Lilliana was the first target with the 0.4m Cassegrain 
telescope. The curve exhibits 0\fm56 R-filtered amplitude with asymmetric 
humps and unequal minima. Our four observing runs were grouped and the light 
curves acquired during different sessions showed noticeable deformation.
The synodic period turned out to be 9\fh361$\pm0\fh002$. On the web site of Raoul Behrend one can find a different rotation period of 6\fh151 with an amplitude reaching 1\fm0. That result is based on data points obtained by several observers and stamped as temporary. The analysis of our data consisting of four observing nights gives the above mentioned synodic period as lowest-sigma case that is 9\fh361.
When applying 6\fh151 as synodic period our data did not fit as composite lightcurve.
However we cannot exclude the existence of different true value therefore further observation is recommended in order to solve this ambiguity.
The composite lightcurve is shown in Fig. \ref{lillianacomp}.

{\bf 894 Erda}: Being a minor planet with ambiguous behaviour in our 
photometric programme, 894 Erda showed cyclic light variations with small 
amplitude during the observations, as plotted in Fig. \ref{erda}. Interesting 
feature of the curve is the sudden sharp dimming visible near the center of 
the graph. One possible explanation of this phenomenon is the probable 
binarity nature of 894 Erda. Assuming this, the fading was caused by a 
hypothetical secondary component. Unfortunately, we did not succeed in 
observing other eclipses during the following nights.

In addition to our results, 894 Erda was observed during five nights between 
15th and 19th of July, 2001 from Santana Observatory, California, USA 
(Stephens 2002). However those observing sessions are fairly close in time to our presented ones, no dimming events were observed.
This author has derived a rotational period of 4\fh69$\pm0\fh01$ . The 
observed 0\fm08 amplitude is consistent with our data. Our composite lightcurve from three nights data with applying 4\fh69 as rotational period is shown in Fig. \ref{erdacomp}.

{\bf 1026 Ingrid}: This very small asteroid was the faintest investigated 
minor planet by us and a serendipitouos by-product while observing 2448 
Sholokhov: the two previously photometrically uninvestigated asteroids were in 
the same CCD frames on February 18th, 2003. The lightcurve of this 17\fm0 
object presented in Fig. \ref{ingrid} implies fast rotation. Even though the single lightcurve covers less than the full rotation cycle, the time between the two minima is almost certainly very close to half the full rotation period, because the amplitude is so large, $>$0\fm5. Thus, the full period is about 5\fh3$\pm0\fh3$.

{\bf 1108 Demeter}: This mid-sized asteroid exhibited very ambiguous light 
variations on three nights. One of the lightcurves is presented in Fig. 
\ref{demeter}. The R-filtered amplitude is 0\fm15. The bizarre lightcurve is also noticed on the web site of Behrend. The complex light variation of this asteroid is hard to be explained. A possibility can be the fast precession of the rotation axis or binary nature.

In addition to the observations of this asteroid, we have discovered the light 
variability of GSC 2127-0056, which served as a comparison star on August 1, 
2001. One night of data suggests the possible eclipsing binary nature of the 
star (Sz\'ekely, in preparation).

Besides our observations, 151 data points were acquired in Santana Observatory
during seven runs between 9th and 29th of July, 2001 (Stephens 2002), 
resulting in a synodical period of 9\fh70$\pm0\fh01$. The unfiltered amplitude 
was 0\fm17, in good agreement with our results. The published lightcurve is fairly normal with no anomalous features. When applying 9\fh70 as synodical period our composite lightcurve does not fit well as can be seen in Fig. \ref{demetercomp}.

{\bf 1170 Siva}: The lightcurve of 1170 Siva is plotted in Fig. \ref{siva}. 
The complex lightcurve shape reveals the presence of two unequal extrema, however, the observing time-span is rather short and the R-filtered amplitude is only 0\fm07. Based on this lightcurve the synodic period may be about 3\fh5. The Behrend web site gives a period of 5\fh215 with a lightcurve amplitude of only 0\fm04. The reliability of that period determination is low. It can be noted that the ratio of the two mentioned period is almost exactly 2/3. Further observation is recommended for this object.

{\bf 1270 Datura}: 
The asteroid was observed during two nights in 1990 and 1991 (Wisniewski et al. 1997). The detected amplitude was 0\fm41 and the derived synodic period was 3\fh2$\pm0\fh1$. The latter run covered only a fraction of the rotational phase. This, and the two months separation between of the two datasets resulted in a 
large uncertaintity of the rotational period. Our observations suggest an 
amplitude over 0\fm6 and 3\fh4$\pm0\fh3$ synodical period, respectively.
However the two period are consistent within uncertainties.
The R-filtered lightcurve is presented in Fig. \ref{datura}. The plot shows humps with different maxima and minima.

{\bf 1286 Banachiewicza}: Our one short lightcurve of 1286 Banachiewicza, covers a minimum and shows an amplitude of variation greater than 0\fm4. Behrend lists a period of 8\fh631 for the period and an amplitude of 0\fm54. Our short dataset of only 1\fh2 duration is consistent with these parameters.

{\bf 1400 Tirela}: The phase diagram of 1400 Tirela using data of four 
observing nights is plotted in Fig. \ref{tirelacomp}. It is worth mentioning that this composite lightcurve does not cover fully the rotational period.
There are noticeable 
irregularities on both branches likely to be caused by surface structures. The 
amplitude of variations is 0\fm55. The rotational period is 
13\fh356$\pm0\fh007$. On the web site of Behrend is a single-night lightcurve covering less than a full cycle with amplitude of 0\fm3. Behrend suggests a period of 10\fh3$\pm$5\fh8, which spans a range consistent with our derived period. In spite of the fact that our result has higher reliability, we cannot exclude the possibility of another period. Further observations are recommended for this small asteroid.

{\bf 1503 Kuopio}: The composite lightcurve of 1503 Kuopio plotted in Fig. 
\ref{kuopiocomp} shows asymmetric humps with strikingly even branches and 
sharp minima. Individual R-filtered lightcurves have low scatter and exhibit 
an amplitude of 0\fm77. The obtained synodic period is 9\fh98$\pm0\fh03$.
Behrend gives 9\fh958 as synodic period calculated from fully covered rotational phase thus the period reliability is rather high. The observed amplitude varied between 0\fm69 and 1\fm01.

{\bf 1506 Xosa}: One of the obtained lightcurves of 1506 Xosa is given in 
Fig. \ref{xosa}. The plot displays light variations with R-filtered amplitude 
of 0\fm22. Though there are signs of periodicity shown here, our data were 
insufficient to calculate rotational period.

Robinson \& Warner (2002) also observed this asteroid, concluding that the 
synodical period was 5\fh9$\pm0\fh01$. Their observations revealed the 
amplitude to be about 0\fm28, which agrees with our result. Our noise hampered datasets composited with 5\fh9 as synodical period does not fit well as can be seen in Fig. \ref{xosacomp}.

{\bf 1695 Walbeck}: The lightcurve of 1695 Walbeck is shown in Fig. 
\ref{walbeck}. We found ambiguous light variations during 3 observing sessions with amplitude exceeding 
0\fm34. In spite of the large amplitude, we were not able to determine the 
period of variations, however it can be implied around 5\fh3. There is no other lightcurve in the literature.

{\bf 2448 Sholokhov}: Spanning just 3\fh6, the lightcurve of 2448 Sholokhov 
exhibited steadily increasing brightness. Unfiltered amplitude is above 
0\fm25, and the rotation period appears to be longer than 14 hours, probably in the range 20-24 hours. We found no other lightcurve in the literature.

{\bf 3682 Welther}: The lightcurve of 3682 Welther, acquired on August 30, 
2001, showed fast
variations. The curve plotted in Fig. \ref{welther} exhibits sharp dimmings 
with different minima
and unequal maxima. The observed unfiltered amplitude was 0\fm35. The shape of 
the lightcurve
showed noticeable changes in merely a two-days long interval, as can be seen 
in the presented
composite lightcurves in Fig. \ref{welthercomp} (especially between 0.8 and 1.0 
phase in the bottom
panel).

This is similar what has been found by Stephens et al. (2002).
In the course of a collaborative observing campaign (Collaborative Asteroid 
Lightcurve Link, CALL)\footnote{\tt 
http://www.minorplanetobserver.com/astlc/default.htm}, they acquired data for 
this minor planet on 36 nights in the autumn of 2001. Their run was initiated 
by an unexpected brightness variation attributed to the probable binarity of 
the asteroid. However, after a detailed analysis of the extensive dataset, the 
authors assigned the intensity drop (accompanied with two other suspicious 
events) to an artifact of the reduction procedure, i.e. the asteroid moved 
close to relatively bright stars that likely affected the background level 
determination. We have checked this possibility for our observations and found 
that there was indeed a 7 mag star just 4 arcminutes from the asteroid on Oct. 
8, 2001, when the lightcurve suggested a rapid brightening. Since we have 
used the Schmidt telescope of the Konkoly Observatory, it might be possible 
that some reflection within the telescope caused such an 
artifact as that of Stephens et al. (2002). Nevertheless, we cannot firmly 
exclude the 
possibility of real intrinsic changes of the asteroid itself. The derived 
synodic period is 3\fh597$\pm0\fh001$ and it is in agreement with 
3\fh5973$\pm0\fh0003$ as given by Stephens et al. (2002). 

{\bf 5349 Paulharris}: Taken under very unfavourable weather conditions, the 
R-filtered photometric data of the small-sized 5349 Paulharris showed only a slight monotonic increase in brightness of about 0\fm06 in 0\fh7. If the trend in brightness is real, it implies a period of greater than 3 hours.
Due to this meagre and noise hampered dataset we cannot establish firm rotational parameters. This is the first published lightcurve for this minor planet.

{\bf 5690 1992 EU}: This asteroid was observed on two nights in the autumn of 
2000. The lightcurves resulted in from the sessions are given in Fig. 
\ref{5690}.
The data suggest an amplitude of 0\fm4, while no firm conclusion is drawn on the
period although based on the lightcurve it can be around 5 hours. The is fit is obtained with synodical period as 6\fh457 as can be seen in Fig\ref{5690comp}.  
There is no other lightcurve in the literature.

{\bf 6510 Tarry}: The lightcurve of 6510 Tarry (Fig. \ref{tarry}) shows one 
and a half different humps with unequal maxima. We were not able to fully 
track the second rising-branch, therefore, it is possible that the amplitude 
is larger than the observed one (0\fm54 in R). The synodic period can be quite close to 7 hours. This is the first published lightcurve for this small asteroid.

{\bf 7505 Furushu }: This small asteroid was observed under its temporary designation as 1997 AM$_{\rm 2}$ in autumn of 2000. The composite phase diagram for 7505 Furushu is given in Fig. \ref{7505comp}. This minor planet showed 
high-amplitude variations exceeding 0\fm75. The computed synodic period is 
4\fh14$\pm0\fh035$. 
There are data in the literature referring to synodical period and amplitude 
of this small asteroid (Stephens 2001). Our data are fully consistent with 
their derived rotational period, 4\fh14$\pm0\fh02$ and the observed amplitude, 
0\fm74, respectively.

{\bf 11436 1969 QR}: This small-sized and faint minor planet had fairly 
detectable amplitude of 0\fm27. The computed phase diagram given in Fig. 
\ref{1969qrcomp}.
The synodical period is 2\fh65$\pm0\fh24$. The results are affected by the 
fact that the exposure times were 3 to 5 minutes because of the faintness of 
the asteroid, which are quite a significant fraction of the derived period. So 
that it is likely that our data were slightly undersampled. On the other hand the combination of short period and long exposures may cause the decreasing of amplitude due to averaging of brightness during exposing. There is no other lightcurve in the literature referring to this asteroid.





\vskip1cm
{\bf Summary}
\vskip1cm

We presented CCD R-filtered and unfiltered photometric data for 23 small and 
intermediate-sized main-belt asteroids. 8 of them were not investigated 
previously. 17 minor planets exhibited detectable light variations. With 
composite lightcurves, we derived synodical periods for seven objects. In case of three minor planets we estimated the rotational period. For the rest only 
lower limits of rotational parameters can be concluded. 
In two cases (894 Erda, 3682 Welther), we have found rapid brightness changes 
superimposed on the much slower rotationally induced variations, which might be attributed to possible binarity, though the presented datasets are far shorter than that is required for drawing firm conclusions on such an interesting issue.
Judged from the plot of the synodical period versus asteroid size, it is clearly discernible that smaller asteroids tend to have shorter rotational periods. However, it is rather difficult to say something conclusive about this topic based on this insufficiently large number of target objects.
The photometric properties of the studies minor planets are summarized in Table\ \ref{vege}.

\vskip1cm
{\bf Acknowledgements:}\\
\noindent This work has been supported by the FKFP Grant 0010/2001, 
OTKA Grants \#F043203, \#T042509 and the Australian 
Research Council. The warm hospitality of the staff of the Konkoly Observatory
and their provision of telescope time is gratefully acknowledged.
The NASA ADS Abstract Service was used to access data and references. Sz. P. wishes to thank the very useful comments and suggestions of Dr. Alan. W. Harris.

\clearpage

{\bf References}
\vskip1cm\noindent

Angeli, C. A., Guimar\~{a}es, T. A., Lazzaro, D., Duffard, R., Fernández, S., 
Florczak, M., Mothé-Diniz, T., Carvano, J. M., \& Betzler, A. S., 2001. 
Rotation periods for small Main-Belt Asteroids from CCD photometry. AJ,	121, 
2245-2252

\vskip3mm\noindent
Binzel, R. P., Farinella, P., Zappala, V., \& Cellino, A., 1989. Asteroid 
rotation rates - Distributions and statistics. In: Asteroids II, ed. R.P. 
Binzel, T. Gehrels, \& M.S. Matthews, Univ. Arizona
Press, 416-441

\vskip3mm\noindent   
Fulchignoni, M., Barucci, M. A., Di Martino, M., \& Dotto, E., 1995. On the 
evolution of the asteroid spin. A\&A, 299, 929-932
    
\vskip3mm\noindent
Harris, A. W., 1997, ``Minor Planet Lightcurve Parameters''\\ 
{\tt http://cfa-www.harvard.edu/iau/lists/LightcurveDat.html}

\vskip3mm\noindent
Kiss, L. L., Szab\'o, Gy., \& S\'arneczky, K., 1999. CCD photometry and new 
models of 5 minor planets. A\&AS, 140, 21-28

\vskip3mm\noindent
Morbidelli, A., Jedicke, R., Bottke, W. F., Michel, P., \& Tedesco, E. F., 		
2002. From magnitudes to diameters: The albedo distribution of Near Earth 
Objects and the Earth collision hazard. Icarus, 158, 329-342

\vskip3mm\noindent
Pravec, P., \&  Harris, A. W., 2000. Fast and Slow Rotation of Asteroids. 
Icarus, 148, 12-20

\vskip3mm\noindent
Robinson, L. E., \& Warner, B. D., 2002. A collaborative work on three 
asteroid lightcurves: 506 Marion, 585 Bilkis, 1506 Xosa. The Minor Planet 
Bulletin, 29, 6-7

\vskip3mm\noindent
S\'arneczky, K., Szab\'o, Gy., \& Kiss, L. L., 1999. CCD observations of 11 
faint asteroids. A\&AS, 137, 363-368

\vskip3mm\noindent
Sheridan, E. E., 2002. Rotational periods and lightcurve photometry of 697 
Galilea, 1086 Nata, 2052 Tamriko, 4451 Grieve, and (27973) 1997 TR25. The 
Minor Planet Bulletin, 29, 32-33

\vskip3mm\noindent
Stellingwerf, R. F., 1978. Period determination using phase dispersion 
minimization. ApJ, 224, 953-960

\vskip3mm\noindent
Stephens, R. D., 2001. Rotational periods and lightcurves of 1277 Dolores, 666 
Desdemona and (7505) 1997 AM2. The Minor Planet Bulletin, 28, 28-29

\vskip3mm\noindent
Stephens, R. D., Warner, B., Pravec, P., Kusnirak, P., Sarounova, L., 			Wolf, 
M., \& Malcolm, G., 2002. Lightcurves of 3682 Welther. The Minor Planet 
Bulletin, 29, 41-46

\vskip3mm\noindent
Stephens, R. D., 2002. Photometry of 866 Fatme, 894 Erda, 1108 Demeter, and 
3443 Letsungdao. The Minor Planet Bulletin, 29, 2-3

\vskip3mm\noindent
Szab\'o, Gy. M., Cs\'ak, B., S\'arneczky, K., \& Kiss, L. L., 2001.
Photometric observations of 9 Near-Earth Objects. A\&A, 375, 285-292

\vskip3mm\noindent
Wang, X. B., 2003. Abstracts of Photometry and Polarimetry of Asteroids: Impact on Collaboration, Kharkiv, 32.

\vskip3mm\noindent
Wisniewski, W.Z., Micha\l owski, T.M., Harris, A.W., \& McMillan, R. S., 		
1997. Photometric observations of 125 asteroids. Icarus, 126, 395-449

\clearpage
\begin{center}
{\bf Figure captions}
\end{center}
\vskip1cm
The lightcurve of 469 Argentina.\\
The composite lightcurve of 469 Argentina.\\
The lightcurve of 531 Zerlina.\\
The lightcurves of 546 Herodias.\\
The lightcurve of 549 Jessonda.\\
The lightcurve of 697 Galilea.\\
The composite lightcurve of 756 Lilliana.\\
The lightcurve of 894 Erda.\\
The composite lightcurve of 894 Erda.\\
The lightcurve of 1026 Ingrid.\\
The lightcurve of 1108 Demeter.\\
The composite lightcurve of 1108 Demeter.\\
The lightcurve of 1170 Siva.\\
The lightcurve of 1270 Datura.\\
The lightcurve of 1286 Banachiewicza.\\
The composite lightcurve of 1400 Tirela.\\
The composite lightcurve of 1503 Kuopio.\\
The lightcurve of 1506 Xosa.\\
The composite lightcurve of 1506 Xosa.\\
The lightcurve of 1695 Walbeck.\\
The lightcurve of 2448 Sholokhov.\\
The lightcurve of 3682 Welther.\\
The composite lightcurve of 3682 Welther.\\
The lightcurve of 5349 Paulharris.\\
The lightcurve of 5690 1992 EU.\\
The composite lightcurve of 5690 1992 EU.\\
The lightcurve of 6510 Tarry.\\
The composite lightcurve of 7505 Furushu.\\
The composite lightcurve of 11436 1969 QR.\\

\clearpage
\begin{figure}
\begin{center}
\leavevmode
\psfig{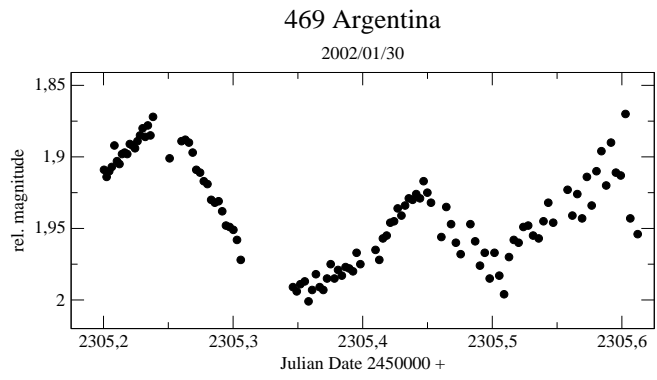}
\caption{The lightcurve of 469 Argentina.}
\label{argentina}
\end{center}
\end{figure}

\begin{figure}
\begin{center}
\leavevmode
\psfig{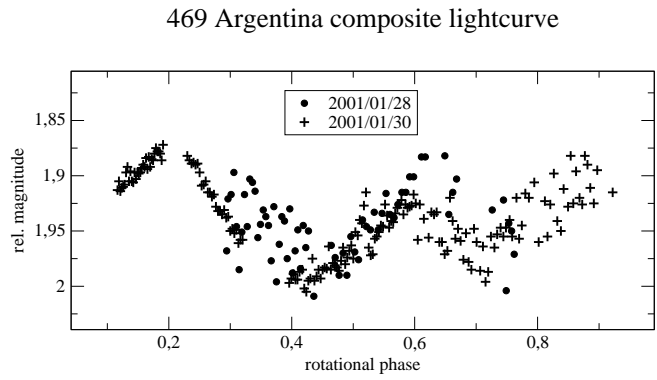}
\caption{The composite lightcurve of 469 Argentina.}
\label{argentinacomp}
\end{center}
\end{figure}

\begin{figure}
\begin{center}
\leavevmode
\psfig{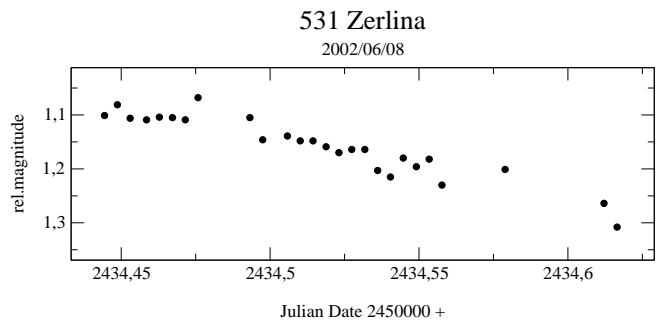}
\caption{The lightcurve of 531 Zerlina.}
\label{zerlina}
\end{center}
\end{figure}

\begin{figure}
\begin{center}
\leavevmode
\psfig{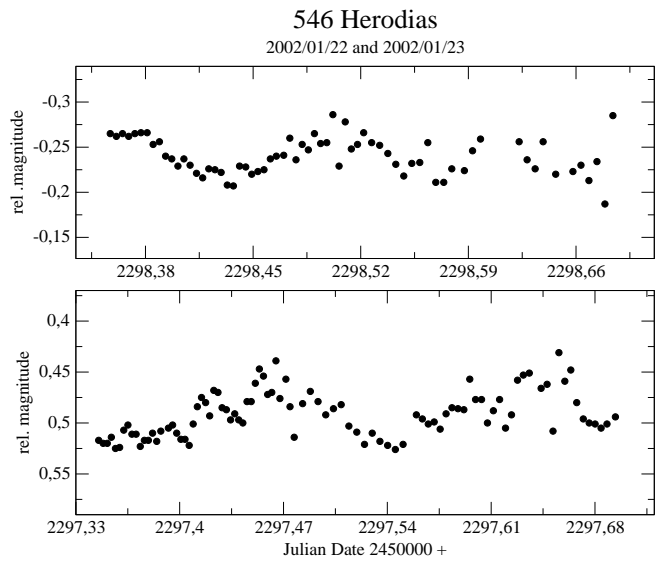}
\caption{The lightcurves of 546 Herodias.}
\label{herodias}
\end{center}
\end{figure}

\begin{figure}
\begin{center}
\leavevmode
\psfig{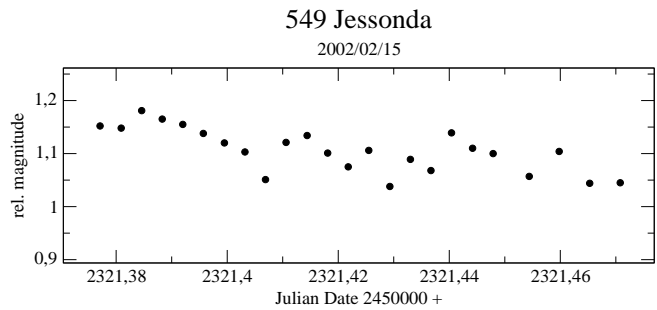}
\caption{The lightcurve of 549 Jessonda.}
\label{jessonda}
\end{center}
\end{figure}

\begin{figure}
\begin{center}
\leavevmode
\psfig{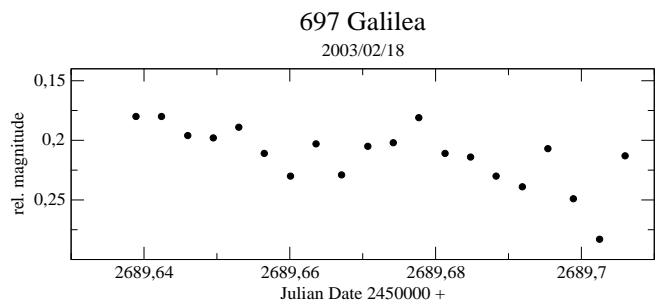}
\caption{The lightcurve of 697 Galilea.}
\label{galilea}
\end{center}
\end{figure}

\begin{figure}
\begin{center}
\leavevmode
\psfig{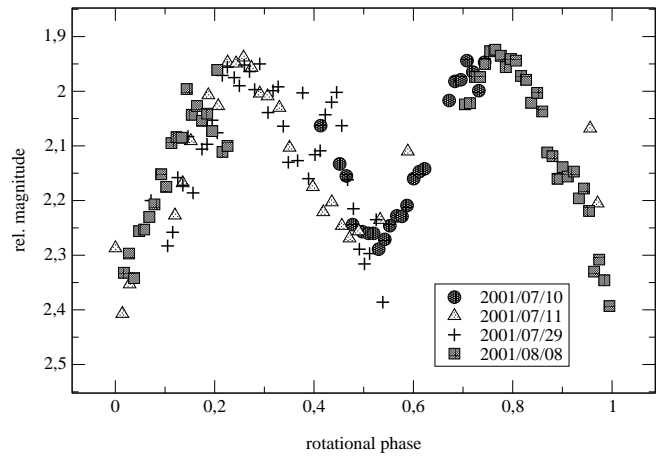}
\caption{The composite lightcurve of 756 Lilliana.}
\label{lillianacomp}
\end{center}
\end{figure}

\begin{figure}
\begin{center}
\leavevmode
\psfig{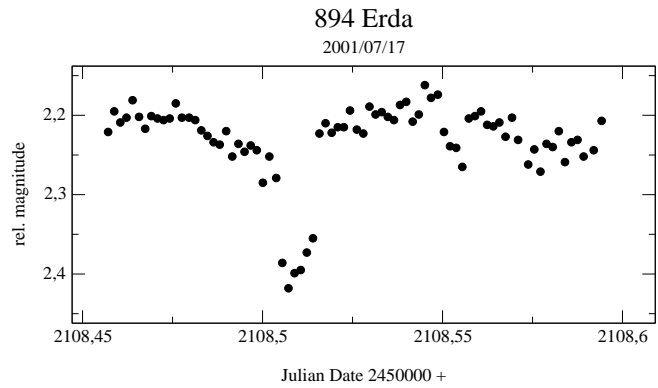}
\caption{The lightcurve of 894 Erda.}
\label{erda}
\end{center}
\end{figure}

\begin{figure}
\begin{center}
\leavevmode
\psfig{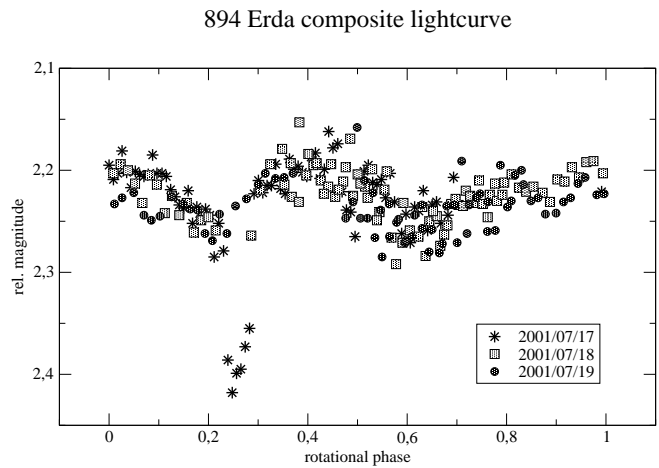}
\caption{The composite lightcurve of 894 Erda.}
\label{erdacomp}
\end{center}
\end{figure}

\begin{figure}
\begin{center}
\leavevmode
\psfig{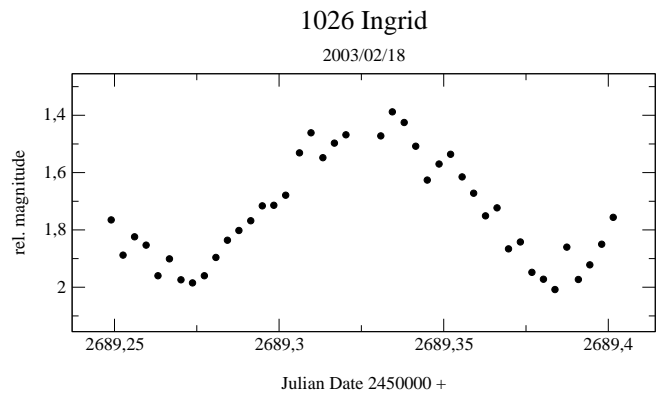}
\caption{The lightcurve of 1026 Ingrid.}
\label{ingrid}
\end{center}
\end{figure}

\begin{figure}
\begin{center}
\leavevmode
\psfig{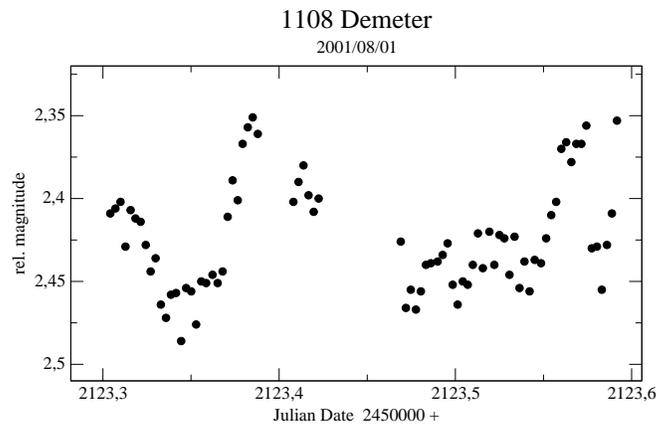}
\caption{The lightcurve of 1108 Demeter.}
\label{demeter}
\end{center}
\end{figure}

\begin{figure}
\begin{center}
\leavevmode
\psfig{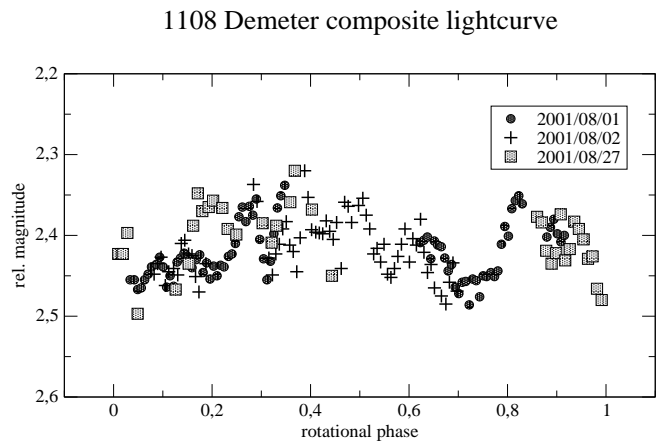}
\caption{The composite lightcurve of 1108 Demeter.}
\label{demetercomp}
\end{center}
\end{figure}

\begin{figure}
\begin{center}
\leavevmode
\psfig{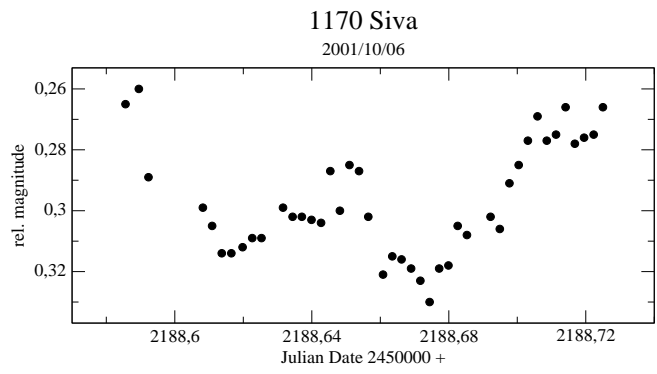}
\caption{The lightcurve of 1170 Siva.}
\label{siva}
\end{center}
\end{figure}

\begin{figure}
\begin{center}
\leavevmode
\psfig{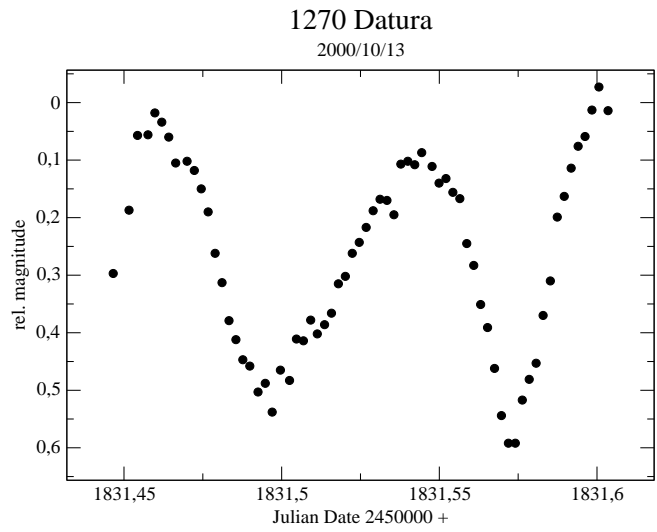}
\caption{The lightcurve of 1270 Datura.}
\label{datura}
\end{center}
\end{figure}

\begin{figure}
\begin{center}
\leavevmode
\psfig{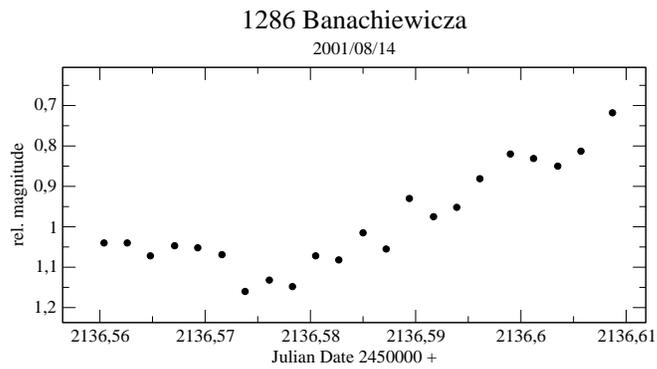}
\caption{The lightcurve of 1286 Banachiewicza.}
\label{banachiewicza}
\end{center}
\end{figure}

\begin{figure}
\begin{center}
\leavevmode
\psfig{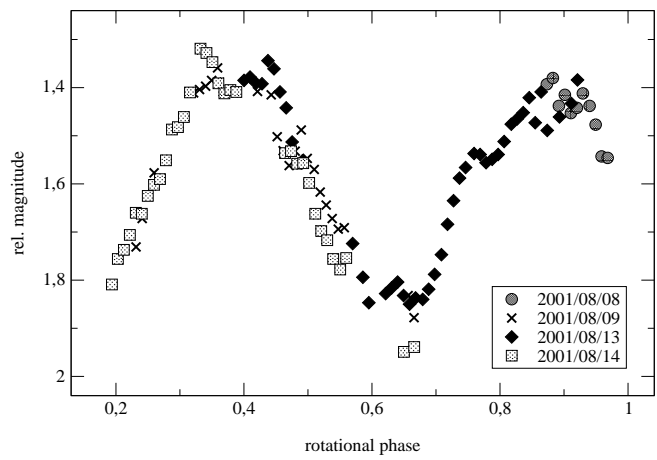}
\caption{The composite lightcurve of 1400 Tirela.}
\label{tirelacomp}
\end{center}
\end{figure}

\begin{figure}
\begin{center}
\leavevmode
\psfig{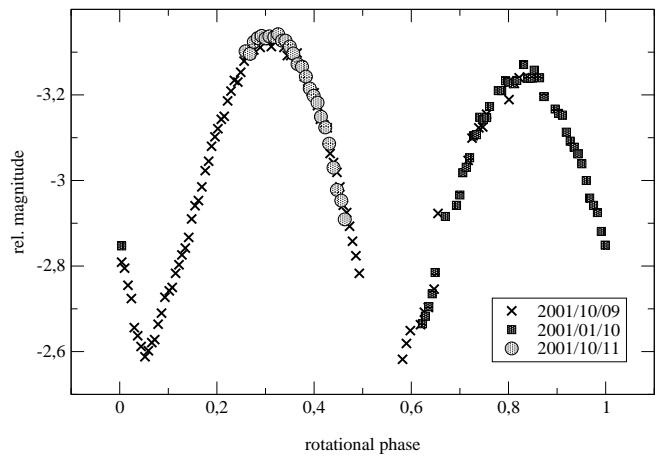}
\caption{The composite lightcurve of 1503 Kuopio.}
\label{kuopiocomp}
\end{center}
\end{figure}

\begin{figure}
\begin{center}
\leavevmode
\psfig{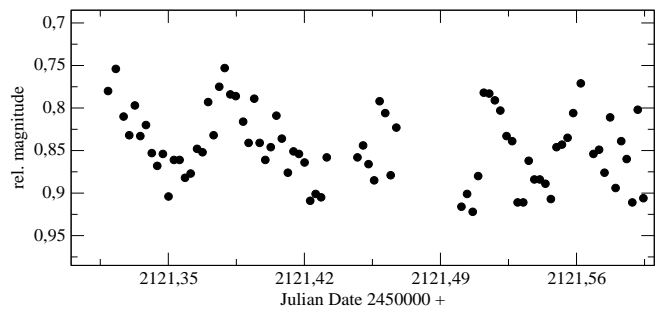}
\caption{The lightcurve of 1506 Xosa.}
\label{xosa}
\end{center}
\end{figure}

\clearpage
\begin{figure}
\begin{center}
\leavevmode
\psfig{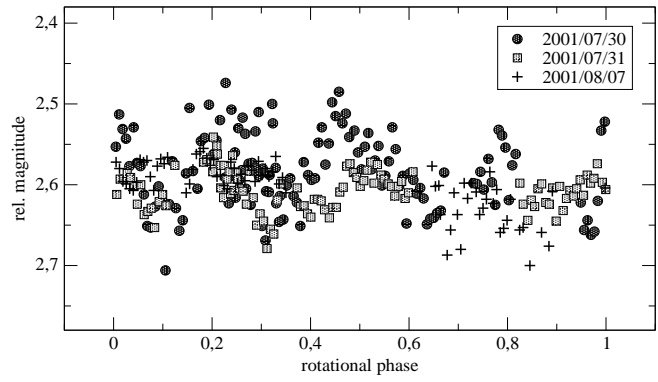}
\caption{The composite lightcurve of 1506 Xosa.}
\label{xosacomp}
\end{center}
\end{figure}

\begin{figure}
\begin{center}
\leavevmode
\psfig{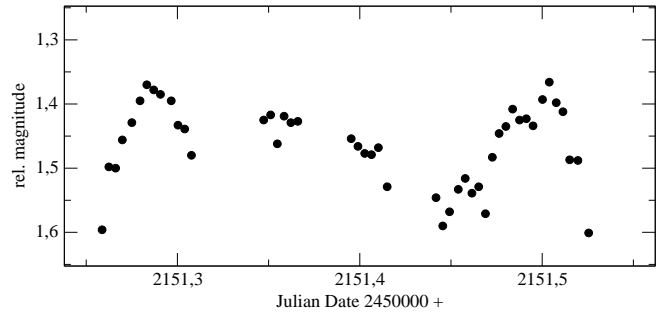}
\caption{The lightcurve of 1695 Walbeck.}
\label{walbeck}
\end{center}
\end{figure}

\begin{figure}
\begin{center}
\leavevmode
\psfig{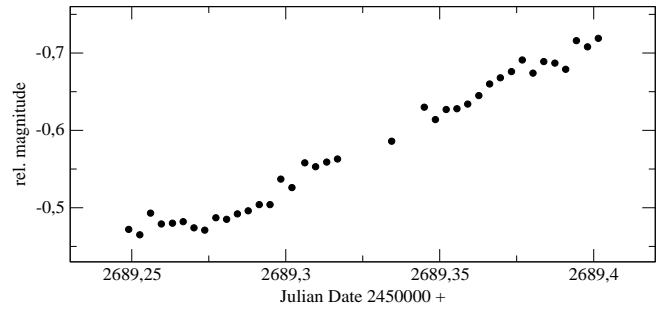}
\caption{The lightcurve of 2448 Sholokhov.}
\label{sholokhov}
\end{center}
\end{figure}

\begin{figure}
\begin{center}
\leavevmode
\psfig{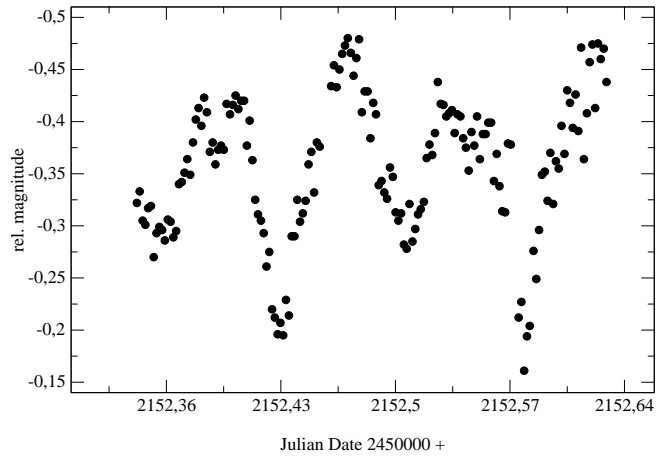}
\caption{The lightcurve of 3682 Welther.}
\label{welther}
\end{center}
\end{figure}

\begin{figure}
\begin{center}
\leavevmode
\psfig{figure=3682comp1.eps,width=8.6cm}
\psfig{figure=3682comp2.eps,width=8.6cm}
\caption{The composite lightcurve of 3682 Welther.}
\label{welthercomp}
\end{center}
\end{figure}

\begin{figure}
\begin{center}
\leavevmode
\psfig{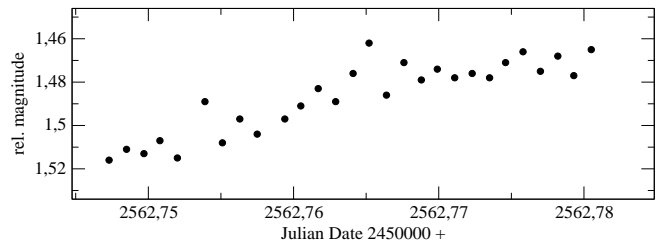}
\caption{The lightcurve of 5349 Paulharris.}
\label{paulharris}
\end{center}
\end{figure}

\begin{figure}
\begin{center}
\leavevmode
\psfig{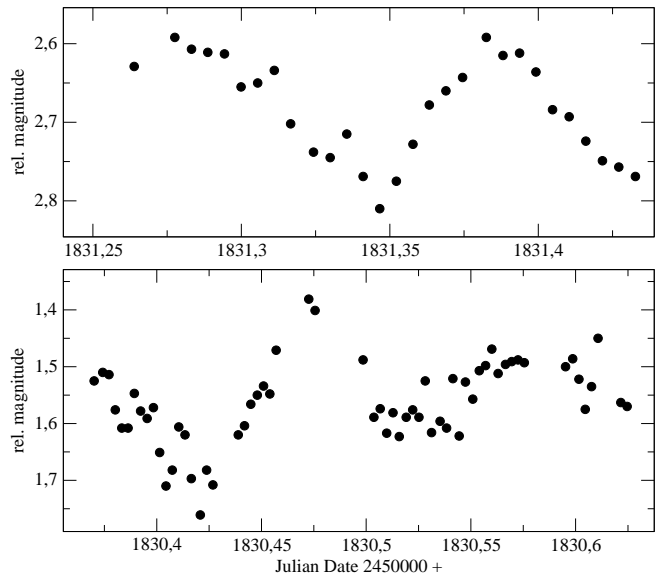}
\caption{The lightcurve of 5690 1992 EU.}
\label{5690}
\end{center}
\end{figure}

\begin{figure}
\begin{center}
\leavevmode
\psfig{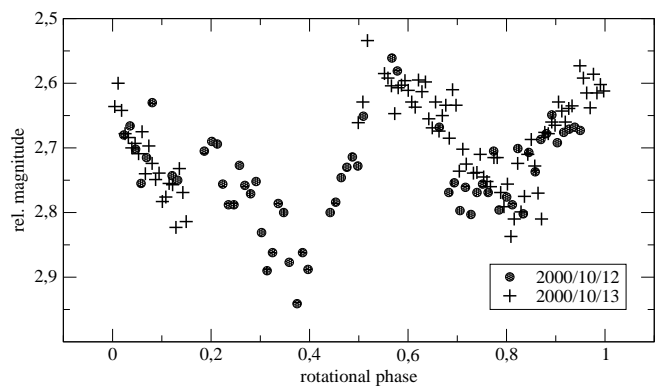}
\caption{The composite lightcurve of 5690 1992 EU.}
\label{5690comp}
\end{center}
\end{figure}

\begin{figure}
\begin{center}
\leavevmode
\psfig{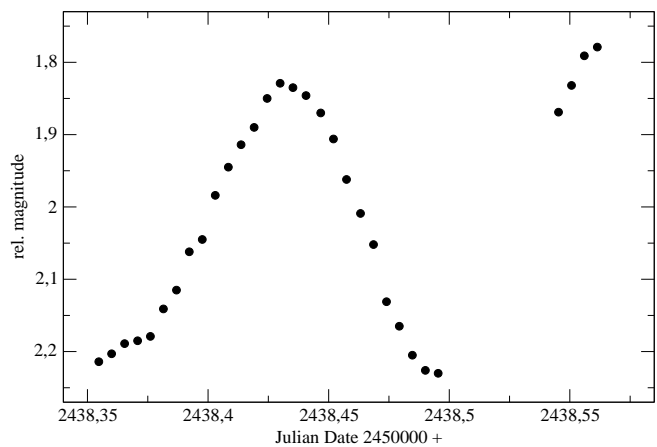}
\caption{The lightcurve of 6510 Tarry.}
\label{tarry}
\end{center}
\end{figure}

\begin{figure}
\begin{center}
\leavevmode
\psfig{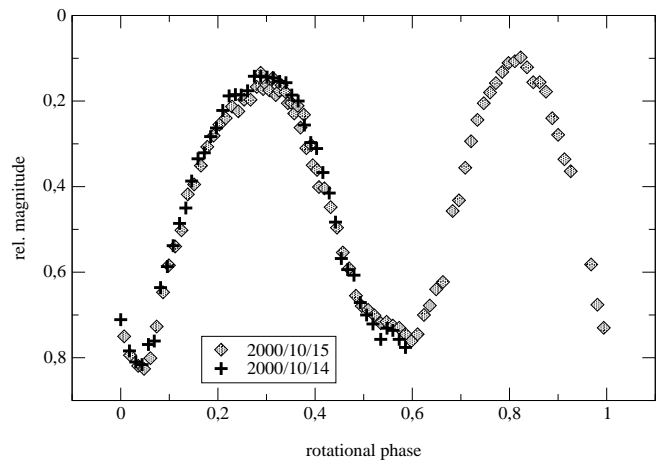}
\caption{The composite lightcurve of 7505 Furushu}
\label{7505comp}
\end{center}
\end{figure}

\begin{figure}
\begin{center}
\leavevmode
\psfig{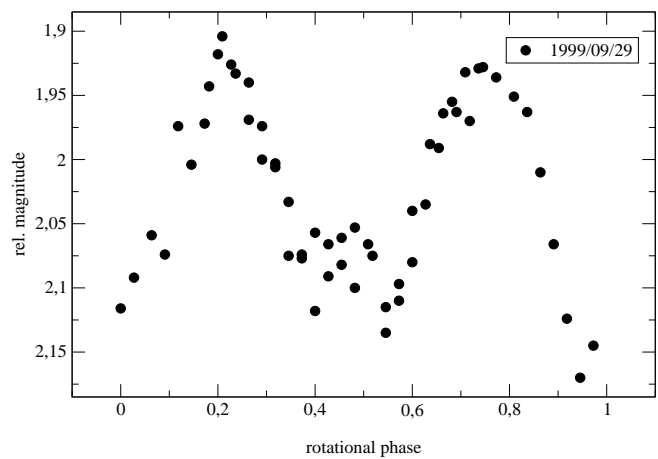}
\caption{The composite lightcurve of 11436 1969 QR.}
\label{1969qrcomp}
\end{center}
\end{figure}

\clearpage

\begin{table}
\caption{Aspect data for the mid-time of observations. Abbreviations: 
r=heliocentric distance [AU], $\Delta$=geocentric distance [AU], 
$\alpha$=solar phase [deg], $\lambda$=geocentric longitude [deg], 
$\beta$=geocentric latitude [deg]}
\begin{small}
\begin{center}
\begin{tabular} {||l||c|c|c|c|c|c|c|c||}
\hline
Asteroid & V & Date$^a$ &r &$\Delta$ &$\alpha$ &$\lambda$ &$\beta$ \\
\hline \hline

469 Argentina
 &12\fm9 & 2002/01/30.91 &2.857 &1.922 &7.69 &110.88 &10.51 \\
531 Zerlina
 & 15\fm4 & 2002/06/08.87 &2.232 &1.445 &20.50 &105.07 &50.46 \\
546 Herodias
 &13\fm2 & 2002/01/23.04 &2.320 &1.449 &14.40 &95.30 &24.00 \\
549 Jessonda
 &13\fm6 & 2002/02/16.04 &2.115 &1.169 &10.45 &124.65 &-2.84 \\
697 Galilea
 &14\fm4  & 2003/02/18.08 &3.287 &2.325 &4.72 &149.30 &15.69 \\
756 Lilliana
 &14\fm2 & 2001/07/12.08 &2.950 &2.042 &10.74 &86.17 &28.64 \\
894 Erda
 &13\fm4 & 2001/07/18.91 &2.741 &1.771 &7.54 &52.18 &18.37 \\
1026 Ingrid
 &17\fm0  & 2003/02/18.91 &2.662 &1.694 &5.43 &136.14 &5.06 \\
1108 Demeter
 &14\fm3 & 2001/02/08.83 &1.804 &0.974 &25.51 &66.02 &47.62 \\
1170 Siva
 &13\fm5 & 2001/07/10.04 &1.744 &0.754 &7.04 &14.76 &12.32 \\
1270 Datura
 &15\fm8 & 2000/10/13.91 &1.955 &1.342 &28.17 &88.23 &-5.37 \\
1286 Banachiewicza
 &14\fm7 & 2001/08/15.04 &2.756 &1.760 &4.71 &36.96 &14.14 \\
1400 Tirela
 &15\fm0 & 2001/08/09.91 &2.377 &1.465 &13.64 &65.32 &25.36 \\
1503 Kuopio
 &14\fm5 & 2001/10/10.91 &2.738 &1.769 &6.21 &11.30 &16.13  \\
1506 Xosa
 &13\fm8 & 2001/07/31.91 &1.936 &0.971 &13.14 &61.44 &23.93 \\
1695 Walbeck
 &15\fm3 & 2001/08/29.79 &2.071 &1.195 &18.30 &52.75 &28.97 \\
2448 Sholokhov
 &14\fm0  & 2003/02/18.91 &2.612 &1.644 &5.54 &136.21 &5.05 \\
3682 Welther
 &13\fm7 & 2001/08/29.91 &1.930 &0.988 &14.97 &10.93 &27.32 \\
3682 Welther
 &13\fm9 & 2001/10/06.91 &1.888 &1.003 &20.05 &17.01 &25.19 \\
5349 Paulharris
 &  14\fm2 & 2002/10/16.08 &1.721 &0.775 &15.60 &27.72 &27.14 \\
5690 1992 EU
 &15\fm6 & 2000/10/12.95 &1.776 &1.218 &32.77 &89.85 &36.29 \\
6510 Tarry
 &15\fm3 & 2002/06/12.95 &1.853 &1.010 &23.68 &73.61 &42.14 \\
7505 Furushu
 &14\fm3 & 2000/10/14.95 &1.719 &0.968 &29.49 &79.58 &-6.77 \\
11436 1969 QR
 &16\fm9 & 1999/09/29.87 &2.055 &1.077 &8.231 &5.98 &12.20 \\ 
\hline
\end{tabular}
\end{center}
\vskip1mm
\hskip3mm $^a$ mid-time of observation
\end{small}
\label{aspect}
\end{table}

\clearpage

\begin{table}
\begin{center}
\caption{The log of the observations}
\label{logs}
\vskip1mm
\begin{small}
\begin{tabular} {||l||l|c|c|c|c||}
\hline
Asteroid & Date &Telescope$^*$ &Filter &Exp. time [sec]&Length [h]\\
\hline \hline
469 Argentina
&2002/01/28 &a & X &40 &6.0 \\
&2002/01/30 &a & X &40 &9.8 \\
&2002/02/01 &a & X &40 &7.4 \\
\hline
531 Zerlina
&2002/06/08 &b & R &90 &4.3 \\
\hline 
546 Herodias
&2002/01/22 &a & X & 45 &8.6 \\
&2002/01/23 &a & X & 45 &8.2 \\
\hline
549 Jessonda
&2002/02/15 &a & X &30 &2.2 \\
\hline
697 Galilea
&2003/02/18 &b & R &90 &1.7   \\	
\hline
756 Lilliana
&2001/07/10 &a & R &45 &3.1 \\
&2001/07/11 &a & R &45 &6.5 \\
&2001/07/29 &a & R &45 &4.8 \\
&2001/08/08 &a & R &45 &5.0 \\
\hline
894 Erda
&2001/07/17 &a & R &45 &3.4 \\
&2001/07/18 &a & R &45 &7.0 \\
&2001/07/19 &a & R &45 &6.5 \\
\hline
1026 Ingrid
&2003/02/18 &b & X &90 &3.6 \\
\hline
1108 Demeter
&2001/08/01 &a & R &45 & 7.2 \\
&2001/08/02 &a & R &45 &6.0 \\
&2001/08/27 &a & R &45 &6.0 \\
\hline
1170 Siva
&2001/10/06 &b & R &60 &3.6 \\
\hline
1270 Datura
&2000/10/13 &b & R &60 &3.6 \\
\hline
1286 Banachiewicza
&2001/08/14 &a & X & 40 &1.2 \\
\hline
1400 Tirela
&2001/08/08 &a & X &45 &1.4 \\
&2001/08/09 &a & X &45 &7.0 \\
&2001/08/13 &a & X &45 &7.0 \\
&2001/08/14 &a & X &45 &6.5 \\
\hline
1503 Kuopio
&2001/10/09 &b & R &60 &8.4 \\
&2001/10/10 &b & R &60 &3.8 \\
&2001/10/11 &b & R &60 &2.2 \\
\hline
1506 Xosa
&2001/07/30 &a & R &45 &6.7 \\
&2001/07/31 &a & R &45 &6.7 \\
&2001/08/07 &a & R &45 &7.4 \\	   
\hline
1695 Walbeck
&2001/08/28 &a & X &60 &6.5 \\
&2001/08/29 &a & X &60 &6.5 \\
&2001/08/30 &a & X &60 &1.7 \\
\hline
2448 Sholokhov
&2003/02/18 &b & X &90 &3.6\\	   
\hline
3682 Welther
&2001/08/29 &a & X &60 &2.4 \\
&2001/08/30 &a & X &60 &7.0 \\
&2001/10/06 &b & R &60 &3.1 \\
&2001/10/08 &b & R &60 &4.6 \\
\hline
5349 Paulharris
&2002/10/15 &b & R  &90 &0.7\\	   
\hline
5690 1992 EU
&2000/10/12 &b & R &60 &6.2 \\
&2000/10/13 &b & R &60 &4.3 \\
\hline
6510 Tarry
&2002/06/12 &b & R &120 &5.3\\
\hline
7505 Furushu
&2000/10/14 &b & R &60 &2.4 \\
&2000/10/15 &b & R &60 &4.8 \\
\hline
11436 1969 QR
&1999/09/29 &b & X  &240 & 4.1 \\
\hline
\end{tabular}
\end{small}
\end{center}
\vskip1mm
\hskip3mm $^*$Used telescopes: (a) 0.4m Cassegrain \hskip3mm (b) 0.6m Schmidt

\end{table}

\clearpage

\begin{table}[!h]
\begin{center}
\caption{The summary of photometric properties}
\label{vege}
\begin{footnotesize}
\vskip1mm
\begin{tabular} {||l||c|c|c|c|c|c|}
\hline
Asteroid & Diameter [km] & $P_{\rm syn}$ [h] & ampl. &  $P_{\rm syn}$ [h] & ampl. & Ref.\\
     & & this work   &   this work &  &  & \\
\hline \hline
469 Argentina & 129 $\pm$3 & 12.3? &0\fm14 & 3? & -- & (a) \\

531 Zerlina & 17.8$\pm$3.7 & $>$3.84 &$>$0\fm17 & 2.352 & 0\fm12 & (b) \\

546 Herodias & 69.7$\pm$1.5 & 10.4? &0\fm07 & & & \\

549 Jessonda & 20.5$\pm$2.1 & -- & $>$0\fm15 & 4.5 & 0\fm1 & (c) \\

697 Galilea & 82.5$\pm$1.7 & -- & $<$0\fm1 & 16.538$\pm$0.002 & 0\fm28 & (d) \\

756 Lilliana & 78.3$\pm$1.5 & 9.361$\pm$0.002 &0\fm56 & 6.151 & 0\fm9 & (b) \\

894 Erda & 40.8$\pm$1.6 &-- & $<$0\fm1 & 4.69$\pm$0.01 &0\fm08 & (e) \\

1026 Ingrid  & 14.4 & 5.3? & $>$0\fm5 & & &\\

1108 Demeter & 27.4 &-- & 0\fm15 & 9.70$\pm$0.01 &0\fm17 & (e) \\

1170 Siva & 12.3$\pm$0.6 & 3.5? &$<$0\fm1 & 5.215 & 0\fm04 & (b) \\

1270 Datura & 9.5$\pm$0.5 & 3.4$\pm$0.3 &0\fm61 & 3.2$\pm$0.1 &0\fm41 & (f) \\

1286 Banachiewicza & 33.8$\pm$4.6 & $>$1.2 &$>$0\fm4 & 8.631 &  0\fm54 & (b) \\

1400 Tirela & 33.0 &13.356$\pm$0.007 & 0\fm55 & 10.3$\pm$5.8 & 0\fm3 & (b) \\

1503 Kuopio & 23.0$\pm$1.7 & 9.98$\pm$0.03 & 0\fm77 & 9.958 & 0\fm69-1\fm01 & (b) \\

1506 Xosa & 30.1 &-- &0\fm22 & 5.9$\pm$0.01 & 0\fm28 & (g) \\

1695 Walbeck & 21.0$\pm$0.7 & 5.3? & 0\fm34 & & & \\

2448 Sholokhov & 33.2$\pm$3.5 & $>$14 & $>$0\fm25 & & & \\

3682 Welther & 33.0 &3.597$\pm$0.001 & 0\fm35 & 3.5973$\pm$0.0003 & 0\fm31 & (h) \\

5349 Paulharris & 19.0 & $>$3 & $>$0\fm06 & & & \\

5690 1992 EU & 20.8 & 5? & $>$0\fm25 & & & \\

6510 Tarry & 18.2 & 7? & 0\fm54 & & & \\

7505 Furushu & 27.5 & 4.14$\pm$0.035 & 0\fm75 & 4.14$\pm$0.02 &0\fm74 & (i) \\

11436 1969 QR & 7.8 & 2.65$\pm$0.24 & 0\fm27 & & & \\
\hline
\end{tabular}
\vskip2mm
References: (a) Wang (2003); (b) Behrend web site; (c) CALL web site; (d) Sheridan (2002); (e) Stephens (2002); (f) Wisniewski et al. 
(1997);
(g) Robinson \& Warner (2002); (h) Stephens et al. (2002); (i) Stephens (2001).
\end{footnotesize}
\end{center}
\end{table}

\end{large}
\end{document}